\documentclass[reprint, aps]{revtex4-2}
\usepackage[utf8]{inputenc}
\usepackage{graphicx}
\usepackage{amsmath, amssymb}
\usepackage{xcolor}
\usepackage{upgreek}
\usepackage{bm}
\usepackage{hyperref}

\newcommand{\crit}{\mathrm{cr}}
\newcommand{\D}{\mathrm{d}}
\newcommand{\ssb}{\mathbf{s}}
\newcommand{\Eb}{\mathbf{E}}
\newcommand{\Bb}{\mathbf{B}}
\newcommand{\vb}{\mathbf{v}}
\newcommand{\Omegab}{\bm{\Omega}}

\begin{document}
	
	\title{Acceleration of spin-polarized proton beams via two parallel laser pulses}

	\author{Lars Reichwein}
	\email{lars.reichwein@hhu.de}
	\affiliation{Institut f\"{u}r Theoretische Physik I, Heinrich-Heine-Universit\"{a}t D\"{u}sseldorf, 40225 D\"{u}sseldorf, Germany}
	
	\author{Markus B\"{u}scher}
	\affiliation{Peter Gr\"{u}nberg Institut (PGI-6), Forschungszentrum J\"{u}lich, 52425 J\"{u}lich, Germany}
	\affiliation{Institut f\"{u}r Laser- und Plasmaphysk, Heinrich-Heine-Universit\"{a}t D\"{u}sseldorf, 40225 D\"{u}sseldorf, Germany}
	
	\author{Alexander Pukhov}
	\affiliation{Institut f\"{u}r Theoretische Physik I, Heinrich-Heine-Universit\"{a}t D\"{u}sseldorf, 40225 D\"{u}sseldorf, Germany}
	
	\date{\today}
	
	\begin{abstract}
		We present a setup for highly polarized proton beams using two parallel propagating laser pulses that have a carrier envelope phase difference of $\pi$. This mechanism is examined utilizing particle-in-cell simulations and compared to a single-pulse setup commonly used for magnetic vortex acceleration. We find that the use of the dual-pulse setup allows for peak energies of 124 MeV and good angular spread for two pulses with normalized laser vector potential $a_0 = 100$. Compared to a single pulse, we further observe higher polarization of the accelerated bunch.
	\end{abstract}
	
	\maketitle
	\section{Introduction}
	High-energy spin-polarized particle beams can be utilized to examine the nuclear spin structure or even retrieving transient magnetic fields in laser-plasma interaction \cite{Ageev2005, Gong2021}. In recent years, laser-plasma based acceleration schemes of spin-polarized particle beams have gained a lot of interest due to the compactness of the needed setups. This has brought forward the numerical investigation of several acceleration schemes (see \cite{Buescher2020} for a general overview).  A first experimental study on the acceleration of such polarized beams has recently been carried out at \textsc{Phelix} (GSI Darmstadt) using a hyperpolarized $^3$He gas jet target \cite{Fedorets2022}. The data from this measurement are still being analyzed.
	
	In the case of electrons, wakefield acceleration (both laser- and particle beam-driven) has been theoretically shown to deliver high polarization electron bunches if the plasma target is pre-polarized. Generally, such targets are needed since no significant polarization build-up can be achieved during the laser-plasma interaction. There has, however, recently been a publication by Nie \textit{et al}, which showed that a polarization of up to 30\% can be obtained from initially unpolarized targets \cite{Nie2021}. Using pre-polarized targets, a polarization of up to 80\% seems achievable \cite{Wu2019, Wu2019a}. The quality of the beam strongly depends on the laser and target parameters, for example in the case of laser-driven wakefield acceleration it was shown that Laguerre-Gaussian (LG) laser pulses are beneficial as the weaker azimuthal magnetic fields leads to less depolarization during injection of the electrons. This comes at the cost of the beam shape, since the LG driver creates a donut-like electron beam \cite{Wu2019a}.
	
	For protons the options become more sparse: while several schemes usable for the efficient acceleration of proton beams are known, only a few of them allow for the target's pre-polarization. Setups like target normal sheath acceleration would require pre-polarized cryogenic hydrogen foils which are experimentally extremely challenging \cite{Buescher2020}. The same problem arises in the case of multi-layer targets \cite{Gong2020}. Further, currently no option for creating spin-polarized beams from initially unpolarized targets, as in the case of electrons, is known.
	
	Methods that would generally be experimentally feasible within the scope of pre-polarization are proton wakefield acceleration \cite{Shen2007} and magnetic vortex acceleration (MVA) \cite{Nakamura2010, Park2019}. For the former, a significantly higher laser vector potential $a_0$ is required as compared to the case of the much less inert electrons. Some theoretical studies regarding this topic have been published by H\"{u}tzen \textit{et al} \cite{Huetzen2020} and Li \textit{et al} \cite{Li2021}. In contrast, MVA can already occur at currently achievable laser energies, although higher energies would still be required for some of the future applications \cite{Burkardt2010}. In the work of Jin \textit{et al} it has been shown that generally MVA can preserve up to 80\% polarization of the final beam \cite{Jin2020}. The main problem is, however, that for high field strengths the final polarization drops accordingly since the stronger fields induce increased precession of the beam particles. Thus, for higher intensities, alternative approaches need to be considered.
	
	As already examined theoretically and experimentally for proton (or electron) acceleration without the consideration of spin precession, different laser modes might be beneficial for better acceleration \cite{Aurand2016, Chen2008}. Up to now the most common approach has been to irradiate foil-like targets with these different modes, and not an extended gaseous target like it is required for spin polarization.
	These mechanisms either use pre-formed Laguerre-Gaussian or Hermite-Gaussian beams, or two (or more) spatially separated beams. In the latter case, the two laser pulses can either co-propagate \cite{Zhou2014} or they may cross at an angle \cite{Ferri2019}. These multi-beam setups could be realized using a fiber-optic setup as proposed by Mourou \textit{et al} \cite{Mourou2013}.
	
	In this paper we consider the acceleration of spin-polarized proton beams using two parallel co-propagating laser pulses. The two linearly polarized Gaussian beams are separated by some transverse distance and have a carrier envelope phase (CEP) difference of $\pi$. When traversing a spin-polarized target, the pulses form channels similar to magnetic vortex acceleration and eject collimated proton beams. Utilizing particle-in-cell (PIC) simulations we show that the overlapping longitudinal electric fields create an accelerating field beneficial for acceleration that maintains a high spin-polarization even for high field strengths that could be achieved in the future at facilities like ELI and XCELS \cite{Gales2018, Bashinov2014, Hollinger2020}. The simulation results are compared to the case of a single beam propagating through the target and the importance of a density down-ramp at the target's end will be discussed.
	
	\section{PIC simulations}
	For the PIC simulations we use the code \textsc{vlpl} \cite{Pukhov1999, Pukhov2016}.
	The simulation box has a size of $120 \times 80 \times 80 \lambda_L^3$ and moves with the laser pulse. Here, $\lambda_L = 0.8$ {\textmu}m is the laser wavelength. The grid size is $h_x = 0.05 \lambda_L$, $h_y = h_z = 0.25 \lambda_L$ ($x$ being the laser propagation direction), although it has to be noted that the scaling feature of \textsc{vlpl} is used which increases the transverse grid size by 5\% per cell for cells with $|y|,|z| > 20 \lambda_L$. In accordance with the RIP solver \cite{Pukhov2020}, a time step of $\Delta t = h_x / c$ is used.
	
	The laser setup is simulated as two separated laser pulses with linear polarization and CEP difference of $\pi$ with respect to one another. Each of the two pulses has a length of $\tau = 26.7$ fs and a focal spot-size of $w_0 = 4$ {\textmu}m.
	The normalized laser vector potential for both pulses will be varied in the range $a_0 = 25 - 100$. The centers of the laser pulses are separated by $\Delta y = 8$ {\textmu}m. At the start of the simulation ($t=0$ fs), both pulses are placed at $x=-16$ {\textmu}m.
	
	As a plasma target, the pre-polarized HCl gas already proposed in several publications \cite{Wu2019a, Jin2020} is used. The hydrogen and chlorine densities are $n_\mathrm{H} = n_\mathrm{Cl} = 0.0122 n_\crit$, such that the electron density is close to the critical density $n_\crit = 1.7 \times 10^{21}$ cm$^{-3}$. Initially, the hydrogen and chlorine atoms are already ionized to $\mathrm{H}^+$ and $\mathrm{Cl}^{2+}$, respectively. Throughout the simulations we only consider field ionization according to the ADK model \cite{Ammosov1986} and neglect impact ionization. The constant density target has a length of $200$ {\textmu}m and is enclosed by an density up-/down-ramp of $4$ {\textmu}m, respectively. The up-ramp starts at $x = 0$ {\textmu}m. Initially, the target is fully spin-polarized, i.e. all protons have spin $s_y = 1$.
	
	For our simulation only the precession of spin according to the T-BMT equation is considered \cite{Mane2005}. This equation describes the precession of particle spin $\ssb$ via
	\begin{align}
		\frac{\D \ssb}{\D t} = - \Omegab \times \ssb \; ,
	\end{align}
	in dependence of the precession frequency
	\begin{align}
		\Omegab = \frac{qe}{mc} \left[ \Omega_\Bb \Bb - \Omega_\vb \left( \frac{\vb}{c} \cdot \Bb \right) \frac{\vb}{c} - \Omega_\Eb \frac{\vb}{c} \times \Eb \right] \; . \label{eq:pref}
	\end{align}
	Here, $qe$ is the charge of the particle in multiples of the elementary charge $e$, $m$ is the particle's mass and $c$ is the vacuum speed of light. Each of the summands in Eq. (\ref{eq:pref}) containing the electric fields $\Eb$, the magnetic field $\Bb$ and the particle velocity $\vb$ is also comprised of one of the three prefactors
	\begin{align}
		& \Omega_\Bb = a + \frac{1}{\gamma} \; , && \Omega_\vb = \frac{a \gamma}{\gamma + 1} \; , && \Omega_\Eb = a + \frac{1}{1 + \gamma}
	\end{align}
	which depend on the particle's anomalous magnetic moment $a$ and its Lorentz factor $\gamma$. Besides the spin precession, the Stern-Gerlach force \cite{Gerlach1922} and the Sokolov-Ternov effect \cite{Ternov1995} could affect the spin and particle dynamics for certain parameter regimes. A full discussion for the relevance of these effects is given by Thomas \textit{et al} in \cite{Thomas2020}. We will neglect them here and focus on the T-BMT equation.
	
	\begin{figure}
		\centering
		\includegraphics[width=\columnwidth]{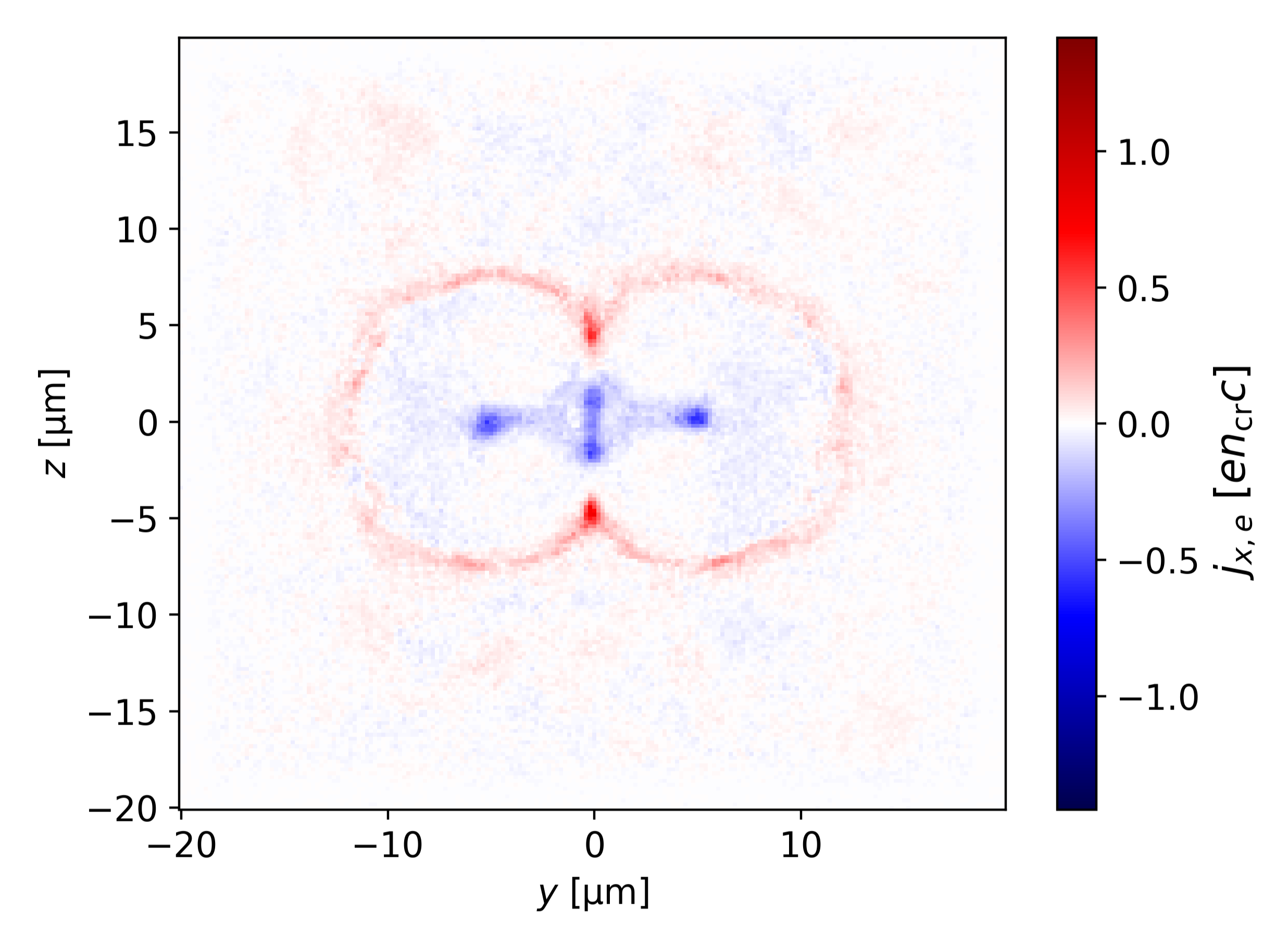}
		\caption{\label{fig:current} Longitudinal currents induced by the electrons in a $y$-$z$-slice of the plasma target when the two channels are formed by the laser pulses at $t= 427$ fs. At the center of each channel and in the region between the two pulses a current (blue) is flowing, which is compensated by a return current at the channel walls (red).}
	\end{figure}
	
	The simulations show that initially the processes in the plasma are quite similar to magnetic vortex acceleration which is explained as follows by Park \textit{et al} in \cite{Park2019}: The laser pulse pushes electrons outwards with its ponderomotive force leaving behind an empty channel. As the laser accelerates electrons in a wake, a thin central filament is formed. This filament carries a strong current, while a corresponding return current is built up in the channel wall, which yields an azimuthal magnetic field inside the channel.
	At the target's end, the transverse expansion of the magnetic field induces a displacement of the electrons and protons, leading to a longitudinal (accelerating) and transverse (focusing) electric field.

\begin{figure*}
		\centering
		\includegraphics[width=\textwidth]{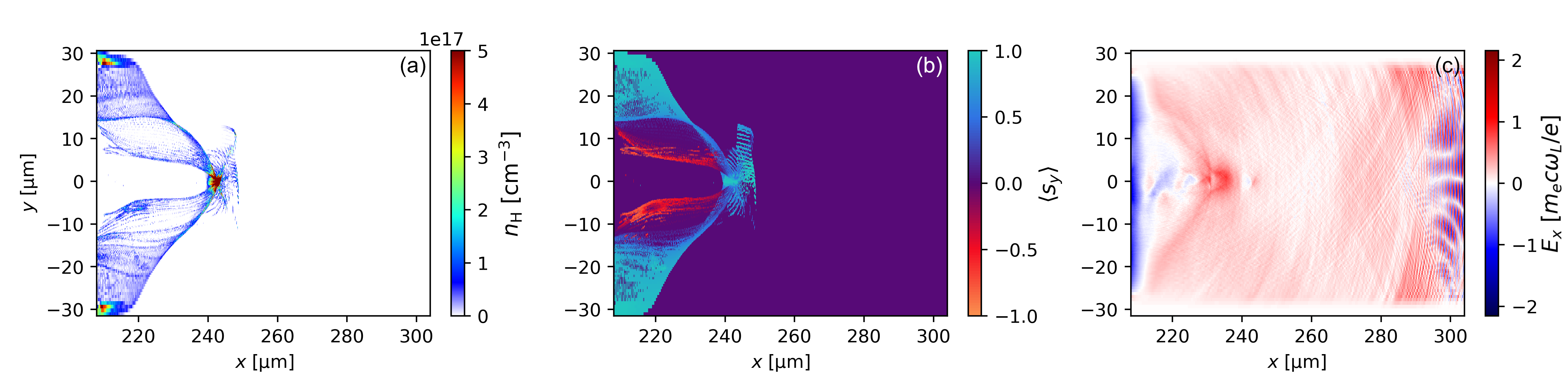}
		\caption{\label{fig:2dframes}(a) Proton density distribution after $t= 1.1$ ps for two $a_0 = 100$ laser pulses propagating through the HCl target. The highest density can be seen in towards the central axis (colorbar clipped for better visibility). (b) Average spin in $y$-direction of the protons. Due to the prevalent electromagnetic fields, part of the protons (see red region around $x = 220$ {\textmu}m) become depolarized. (c) The accelerating field $E_x$ behind the edge of the plasma target.}
	\end{figure*}

	In the case of the dual-pulse scheme, each of the laser pulses creates its own channel. Besides the central filament in each channel, a third filament is formed: as the lasers eject electrons in transverse direction, they form a region of high electron density in the region between the two channels. Now the fact, that the two pulses have a CEP difference of $\pi$ becomes important:
	If the two pulses were polarized in-phase, the overlapping longitudinal electric field $E_x$ would be rather weak and it has been shown that such two channels can attract each other over time \cite{Pukhov1996, Dong2002}. Using the phase difference, however, yields a region of strong accelerating field in the space between the two pulses (although the presence of plasma clearly changes the effective fields compared to the case in vacuum). Further, merging like for in-phase polarization is not possible. This fields drives a current in the third filament (cf. Fig. \ref{fig:current}) and finally an ejection and acceleration process at the end of the target analogous to MVA. The accelerating electric field can also be seen in Fig. \ref{fig:2dframes}.
	
	\section{Results}
\begin{figure}
		\centering
		\includegraphics[width=\columnwidth]{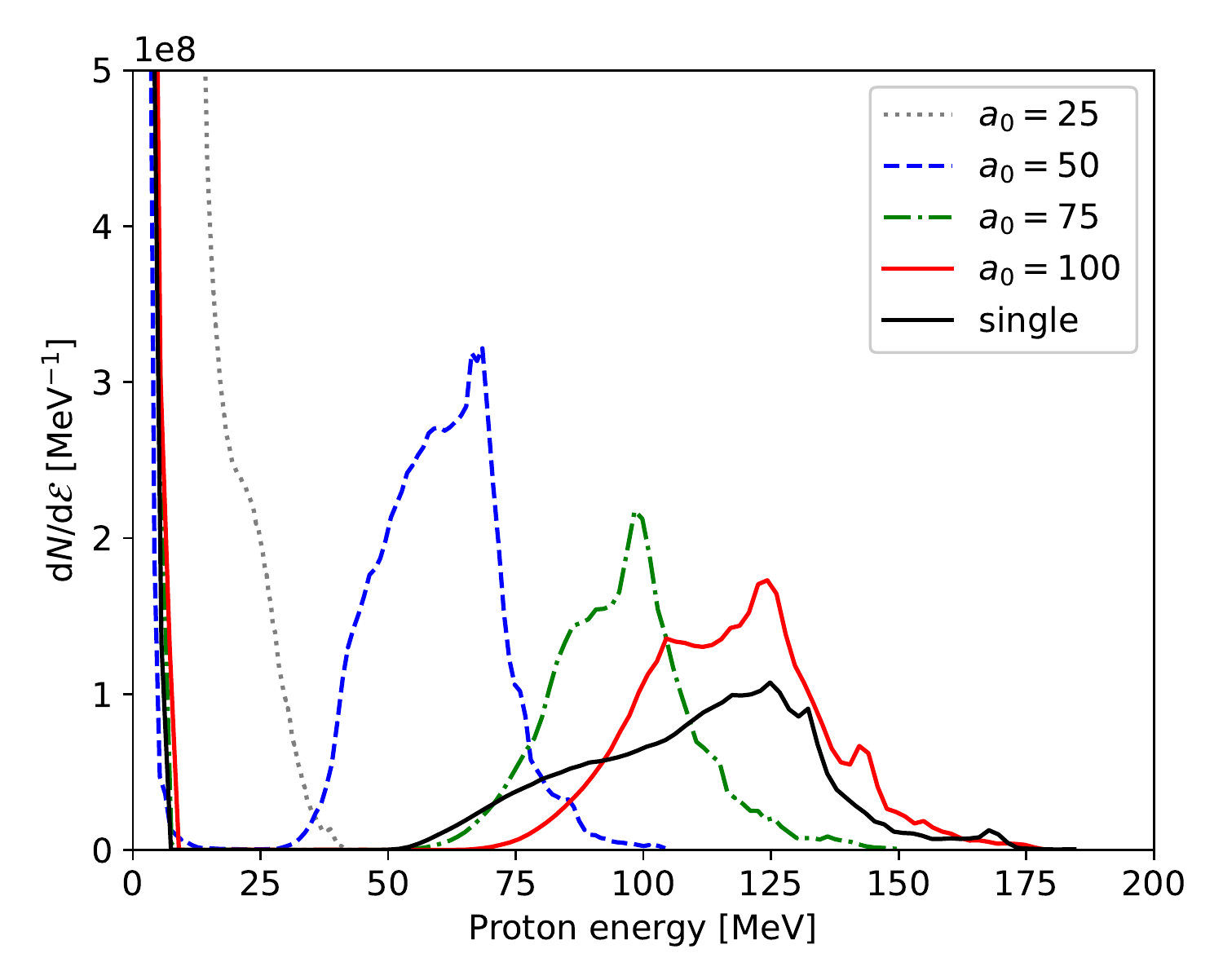}
		\caption{\label{fig:energy_spectrum}Obtained proton energy spectra for different laser vector potentials.}
	\end{figure}

	The results for the different $a_0$ values at $t = 1.3$ ps are summarized in table \ref{tab:data}.  As expected, the maximum obtained energy rises with the laser energy. In the case of $a_0 = 25$ (per laser), a maximum energy of $\mathcal{E}_\mathrm{max} \approx 45 $ MeV is reached, while for $a_0 = 100$ energies of up to 182 MeV are observed. For our further discussion, we will only consider protons with a momentum spread of $\pm 2^\circ$. Looking at Fig. \ref{fig:energy_spectrum} shows that the energy spectra in that case clearly exhibit peaks, which we will refer to as $\mathcal{E}_\mathrm{p}$. One notable exception is the distribution for $a_0 = 25$, which does not exhibit such a peak and therefore will be disregarded from now on. For $a_0=50$, this peak at approx. 68.5 MeV is very narrow, while the distributions for larger $a_0$ appear broader. For $a_0 = 100$, the spectrum seemingly exhibits two smaller peaks around 100 MeV and 150 MeV besides the main peak of 124 MeV. This could potentially indicate that the acceleration mechanism actually consists of several mechanisms.
	
\begin{figure}
		\centering
		\includegraphics[width=\columnwidth]{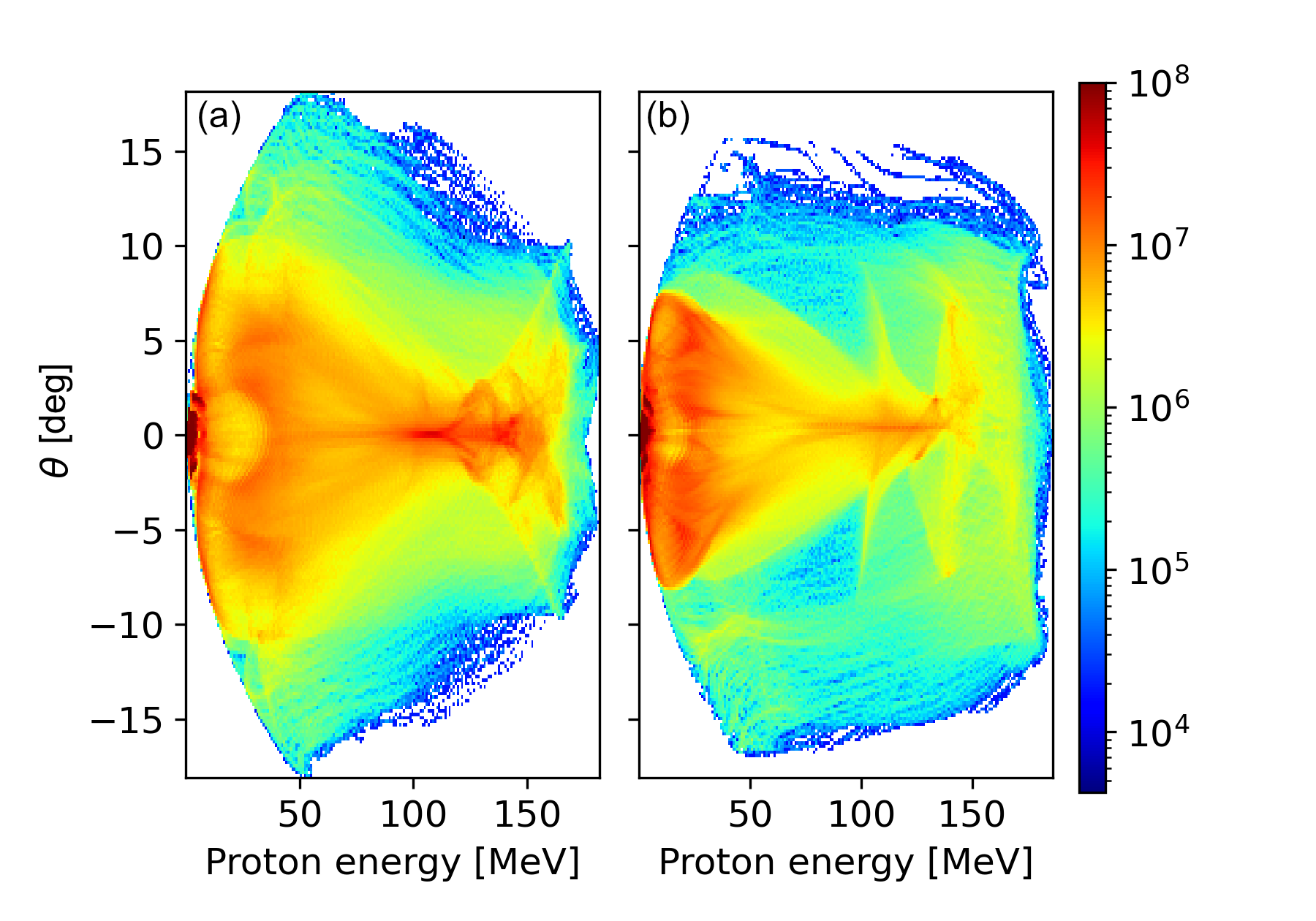}
		\caption{\label{fig:angle}Angular distribution for different energies for the dual-pulse setup with $a_0 = 100$ (a) and the compared single pulse (b). The colormap (clipped at $10^8$) corresponds to the number of particles in that specific angle/energy bin. The angle is defined as $\theta = \arctan(p_y / p_x)$.}
	\end{figure}

	Spatially, the protons are well collimated even for high energies (cf. Fig. \ref{fig:2dframes}). Here, the density for the setup using two $a_0 = 100$ pulses shows a well-defined bunch around $x \approx 240$ {\textmu}m. Depending on the choice of laser vector potential and plasma density, a more shock-like distribution can be observed. Further, besides the central density peak around $y = 0$ {\textmu}m, two weaker density peaks at the two channels' respective centers may be seen (each from a single MVA process). The good angular spread in dependence of energy can further be seen in Fig. \ref{fig:angle}. 
	
	Besides the increase in energy, beam charge is also of interest: for the displayed laser-plasma parameters, charges in the range of (0.61 - 1.07) nC can be obtained from the dual-pulse setup. Note that these numbers refer to the amount of protons in the full width half maximum (FWHM) around the energy peak. A clear trend for the charge is absent from the data, which can partly be attributed that the density is kept the same throughout the different simulations with varying $a_0$. Adapting the target density to the laser strength would likely improve upon the yield. Generally, a higher total amount of protons is accelerated for increased laser energy.
	
	\begin{table}
		\centering
		
		\begin{ruledtabular}
			\begin{tabular}{lcccccc}
				$a_0$ & &$\mathcal{E}_\mathrm{p}$ [MeV] & $\mathcal{E}_\mathrm{max}$ [MeV] & $Q$ [nC] & $P$ [\%] & $\Delta \mathcal{E}/\mathcal{E}_\mathrm{p}$ [\%] \\
				\hline
				50 &(d) & 68.5 & 107.8 & 1.07 & 93 & 40 \\
				75 &(d)& 98.3 & 156.1 & 0.61 &  84 & 25 \\
				100 &(d) & 124.3 & 181.8 & 0.76 & 77 & 29  \\
				\hline
				141 &(s) & 124.8 & 186.3 & 0.61 & 64 & 39
			\end{tabular}
		\end{ruledtabular}
		
		\caption{\label{tab:data} Energy, charge and minimum polarization obtained from the different PIC simulations after $1.3$ ps using different $a_0$ for the dual (d) and single (s) pulse setups. Note that the charge $Q$ and the polarization $P$ refer to the protons in the FWHM around $\mathcal{E}_\mathrm{p}$. The simulation with $a_0 = 25$ is excluded from the table as it does not exhibit a recognizable energy peak.}
	\end{table}
	
\begin{figure}
		\centering
		\includegraphics[width=\columnwidth]{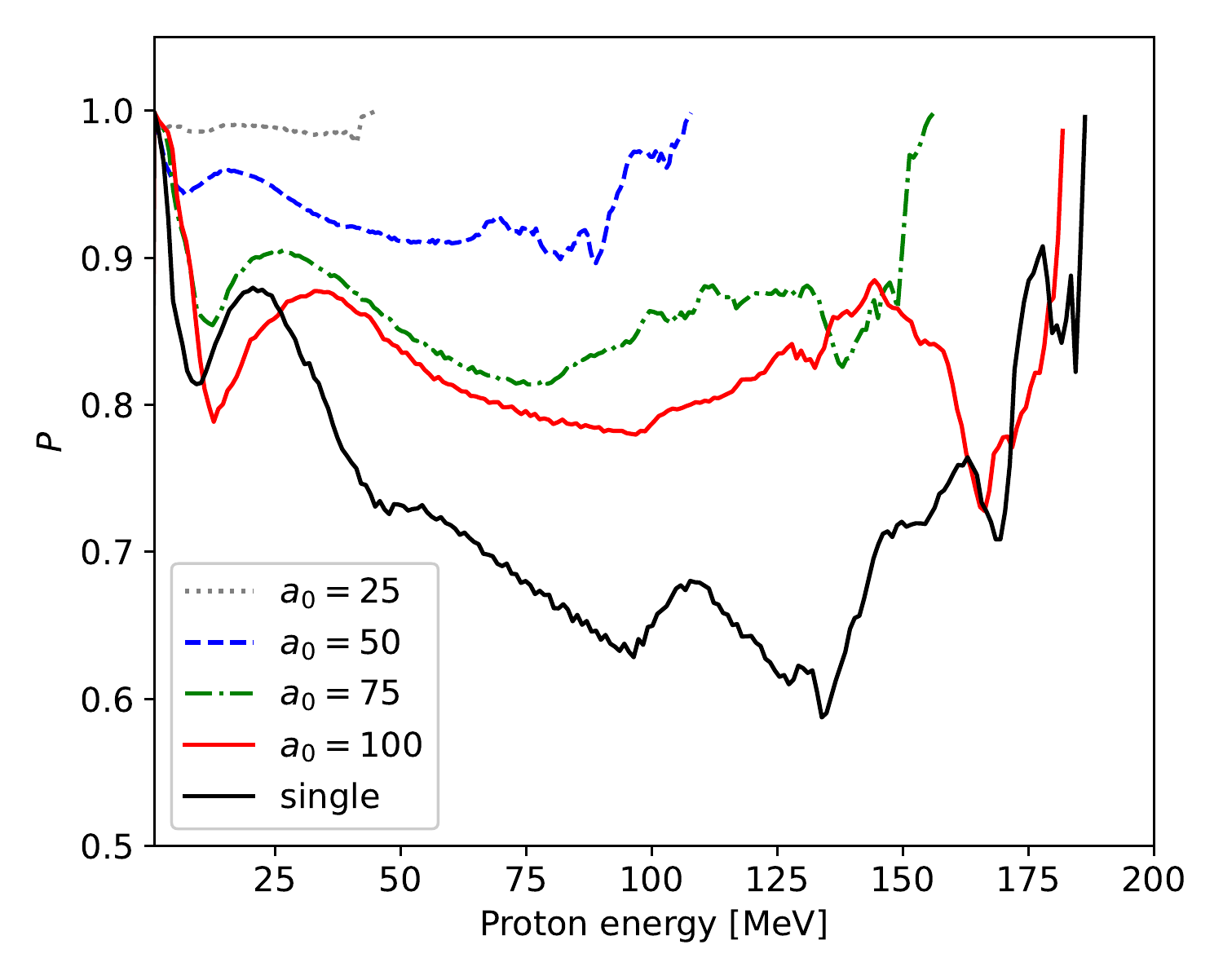}
		\caption{\label{fig:dPdE}Polarization spectra for the different simulations in the whole simulation box. The minimum polarization is only reached for very high energies.}
	\end{figure}

	Finally, the proton bunch's polarization is of concern. The polarization of an $N$-particle ensemble is defined as $P = \sqrt{P_x^2 + P_y^2 + P_z^2}$, where $P_j = \sum_{i} s_{i,j} / N$ is the average over the spin components of all particles in one direction $j \in \lbrace x, y, z \rbrace$.  Figure \ref{fig:2dframes} shows the average $s_y$ value in each cell. Around $x = 220$ {\textmu}m, the spin precesses to a larger extent than in the region of the proton bunch ($x \approx 240$ {\textmu}m). The polarization $P$ in different energy ranges is displayed in Fig. \ref{fig:dPdE}. The general trend is that the polarization decreases as the laser vector potential increases: for $a_0=25$, depolarization is almost negligible throughout the whole energy range.
	
	For the particles in the FWHM around the peak energy, we observe a beam polarization of approx. 93\% at $a_ 0 = 50$ and a drop-off to 77\% for $a_0 = 100$. This is in general accordance with the scaling laws by Thomas \textit{et al} in \cite{Thomas2020} where it was found that the polarization should decrease with rising $a_0$ since the depolarization time scales with $1/ \mathrm{max}(|\Eb|, |\Bb|)$.
	Further, it has to be stressed that for all simulations all polarization spectra exhibit a rising edge for very high energies. This occurs right after reaching the energy range of lowest polarization (see, for example, $\mathcal{E} \geqslant 170$ MeV for $a_0 = 100$) and can be attributed to the lower number of particles in that range and is not a physical result.
	
	\section{Discussion}
	
	The interaction of two laser pulses in a plasma has been examined in various publications \cite{Pukhov1996, Dong2002, Huang2018, Bai2020}, albeit without considering spin polarization. In a publication by Dong \textit{et al}, the interaction of two laser-created channels was studied solving the equations for the pulses co-propagating in the plasma \cite{Dong2002}. This study used the slowly varying envelope of the pulses and neglected the motion of protons that becomes important for the higher field strengths that are considered here. Further, the plasma was assumed to have low density. Still, this publication showed that the two channels can attract each other, an effect which also has been observed in PIC simulations with higher plasma densities \cite{Huang2018, Bai2020}. Simulations using in-phase pulses would display that only lower polarization beams compared to the $\pi$-phase difference would be obtained (not shown here). Other effects that can occur in our present setup can partly be observed in the extensive studies already performed for MVA \cite{Nakamura2010, Park2019}.
	
	\subsection{Comparison to a single-pulse setup}
	For a better of comparison of our scheme with single-pulse setups that are usually used in MVA (like in \cite{Jin2020, Reichwein2021}), we perform a set of simulations with the same target but a single pulse. This linearly polarized Gaussian pulse has the same length and focal spot size but has $\sqrt{2}a_0$ compared to the dual-pulse setup to account for the difference in energy.
	
	These simulations show the typical formation process of the proton bunch already known from MVA. In Figs. \ref{fig:energy_spectrum} and \ref{fig:dPdE} the energy and polarization spectrum for $a_0 = \sqrt{2} \cdot 100$ is shown (black solid line). The peak energy is very similar to the dual-pulse setup, but the spectrum is broader. Further, some higher-energy particles are observed here compared to the dual-pulse scheme. The angular spectrum of the energetic protons is marginally broader, which can be attributed to the fact that the two-pulse setup allow for better collimation of the beam due to stronger fields from both sides.
	
	This is further seen in Figure \ref{fig:angle}, where the angle $\theta = \arctan (p_y / p_x)$ is plotted against the energy. For the dual-pulse setup, the spectrum is symmetric around $\theta = 0^\circ$. In contrast, the single-pulse setup exhibits a slight kink towards positive $\theta$-values for $\mathcal{E} > 100$ MeV. This seems to indicate that the presence of the second laser pulse is able to somewhat suppress instabilities that arise for a single laser pulse. Such a feature is absent in the $x$-$z$-plane. Compared to the dual-pulse setup at $a_0  = 100$ with  $Q \approx 0.76$ nC and $P \approx 77\%$, we observe a lower resulting beam charge of $Q \approx 0.61$ nC and a further decrease of spin polarization to approx. 64\% for the single pulse. The lower polarization could arise from the fact that the protons stem from the channel center in the single-pulse case and are subject to strong electromagnetic fields, whereas the protons from the dual-pulse setup are partly from the region between the two pulses. Therefore, the protons are better shielded from the fields in the latter case, leading to effectively higher polarization.

%
	
	\subsection{Effects of density down-ramps}
	The effects of density-down ramps for MVA were first discussed by Nakamura \textit{et al} in \cite{Nakamura2010}. Due to the decreasing density in the ramp, the electromagnetic fields are able to expand in the direction transverse to propagation. This, in turn, changes the accelerating and focusing fields.
		
	 A specific examination for the influence of down-ramps with respect to spin was performed in \cite{Reichwein2021}. Here it was concluded, that spin polarization is not affected to a great extent in the case of MVA and that mainly the collimation quality of the proton beam suffers for longer ramps.
	 
	 Certainly, for the dual-pulse setup, down-ramps can affect the general dynamics of the system: if the electromagnetic fields expand transversely, the channels might cross each other and diminish the quality of the proton beam. The effects of density modulation for a dual-channel setup were also described in \cite{Dong2002}, but -- as previously mentioned -- with limited applicability to the density and laser parameters used here.
	 
	 As explained in \cite{Reichwein2021}, the spin polarization is expected to remain robust for a variety of ramp lengths. However, when going to experimentally realistic (i.e. longer) targets, the longer interaction time of the protons with the fields will lead to depolarization.

	\section{Conclusion}
	We have shown that a scheme comprising two parallel propagating lasers allows for the acceleration of highly spin-polarized proton beams with approx. 124 MeV peak energy and up to nC charge. Compared to a single pulse setup in typical MVA with the same pre-polarized HCl target, better energy spread and higher polarization are obtained.
	
	Future work will concern other realizable beam profiles for proton acceleration like different Laguerre-Gaussian modes. Further, more analytical work towards spin (de-)polarization in laser-plasma interaction needs to be done. Finally, a setup akin to the one by Nie \textit{et al} \cite{Nie2021} for electrons that achieves spin-polarization for the accelerated particle beam from an initially unpolarized target, would be preferable, as the pre-polarization process complicates the experimental setup.
	
	\begin{acknowledgments}
		This work has been funded in parts by the DFG (project PU 213/9-1). 
		The authors gratefully acknowledge the Gauss Centre for Supercomputing e.V. (www.gauss-centre.eu) for funding this project (qed20) by providing computing time through the John von Neumann Institute for Computing (NIC) on the GCS Supercomputer JUWELS at J\"{u}lich Supercomputing Centre (JSC).  The work of M.B. has been carried out in the framework of the \textit{Ju}SPARC (J\"{u}lich Short-Pulse Particle and Radiation Center, \cite{jusparc}) and has been supported by the ATHENA (Accelerator Technology Helmholtz Infrastructure) consortium.
		L.R. would like to thank X.F. Shen and C. Baumann for helpful discussions over the course of this project.
	\end{acknowledgments}

%

\end{document}